\documentclass[journal,article,submit,pdftex,moreauthors]{Definitions/mdpi} 

\firstpage{1} 
\pubvolume{1}
\issuenum{1}
\articlenumber{0}
\pubyear{2023}
\copyrightyear{2023}
\datereceived{ } 
\daterevised{ } 
\dateaccepted{ } 
\datepublished{ } 
\hreflink{https://doi.org/} 

\Title{Expected Fragment Distribution from the First Interstellar Meteor CNEOS 2014-01-08}

\TitleCitation{Expected Fragment Distribution from the First Interstellar Meteor CNEOS 2014-01-08}

\Author{Amory Tillinghast-Raby $^{\ddagger}$*, Abraham Loeb$^{\ddagger, 1}$ and Amir Siraj $^{\ddagger,2}$}

\AuthorNames{Amory Tillinghast-Raby, Abraham Loeb, and Amir Siraj}

\AuthorCitation{Tillinghast-Raby, A.; Loeb, A.; Siraj, A.}

\address{%
Department of Astronomy, Harvard University, 60 Garden Street, Cambridge, MA 01238, USA \\
$^{1}$ \quad aloeb@cfa.harvard.edu\\
$^{2}$ \quad amir.siraj@cfa.harvard.edu
}

\corres{Correspondence: amory\_tillinghast-raby@alumni.brown.edu}

\secondnote{These authors contributed equally to this work.}

\abstract{In 2014, the fireball of the first interstellar meteor, CNEOS 2014-01-08 (IM1) \citep{SirajarXiv...1904...07224}, was detected  off the northern coast of Papua New Guinea. A recently announced ocean expedition will retrieve any extant fragments by towing a magnetic sled across a 10 km x 10 km area of ocean floor approximately 300 km north of Manus Island \citep{SirajarXiv...2208...00092}. We formulate a model that includes both the probabilistic mass distribution of meteor fragments immediately after the fragmentation event, the ablation of the fragments, and the geographic distribution of post-ablation fragments along the ground track trajectory of the bulk fragment cloud. We apply this model to IM1 to provide a heuristic estimate of the impactor’s post-ablation fragment mass distribution, constructed through a Monte Carlo simulation. We find between $\sim$ 8\% and $\sim$ 21\% of fragments are expected to survive ablation with a mass $\geq$ .001 g, depending on the impactor's empirical yield strength. We also provide an estimation for the geographic distribution of post-ablation fragments.}


\begin{document}

\section{Introduction}

Two interstellar objects have been identified passing through the Solar System over the past five years: first Oumuamua’ in 2017 \citep{2017Nature...552...378}, and then the comet Borisov in 2019 \citep{2020NatAs...4...53}. However, two interstellar meteors (IM) were detected before these reports. CNEOS 2014-01-08 (IM1) was detected in Earth’s atmosphere by U.S Department of Defense (DoD) sensors at 2014-01-08 17:05:34 UTC and was identified as interstellar by \citet{SirajarXiv...1904...07224}. A second interstellar meteor candidate, CNEOS 2017-03-09 (IM2), was identified in September, 2022 \citep{SirajarXiv...2209...09905}. Based on the ram-pressure of the atmosphere at the point where the meteors disintegrated, \citet{SirajarXiv...2209...09905} concluded that both meteors had material strength tougher than iron indicating a source that is unlikely to be a planetary system like the Solar system.

The recovery of fragments from an interstellar object could provide direct material evidence for the chemical composition of its origin \citep{1997JRASC...91...x}. The surface impact location relative to the point in the atmosphere where the meteor explodes is partially dependent on the post-ablation fragment mass because of deceleration due to atmospheric drag. Constructing the post-ablation fragment mass distribution is consequently important for optimizing the search process. The outline of this paper is as follows. In Section 2 we present our mathematical formalism. In Section 3 we summarize the properties of IM1, and in Sections 4 and 5 we apply our model to IM1 and present our results. Finally, we summarize our conclusions in Section 6. 

\section{Mathematical Formulation}
\subsection{Fragmentation model}
Meteoric fragmentation occurs when the ram air pressure acting across the leading face of the impactor exceeds the body's yield strength \citep{2005M&PS...40...817}. After fragmentation, the debris cloud initially moves as a single body with an overall trajectory equal to the flight path of the parent meteor \citep{2005M&PS...40...817}. However, fragments of different sizes experience non-uniform deceleration and,upon slowing to terminal velocity, fall directly downward - differentiating the fragment trajectories as a function of mass.  

A version of the NASA Standard Breakup Model (SBM) modified to describe the atmospheric fragmentation of meteors \citep{2021ICARUS...367...x} provides the pre-ablation fragment mass distribution, defined as $p_{m_{initial}}$,

\begin{equation}
p_{m_{initial}}=\frac{f}{3 \left(D_{\min }^{-f}-D_{\max }^{-f}\right)}\left(\frac{\pi}{6} \rho_m\right)^{f / 3} m_{initial}^{-f / 3-1},  
\end{equation}
where $f = 1.6$, an empirically determined fixed scale factor, $D_{\min}$ is the diameter of the smallest fragment allowed by the distribution, and $D_{\max}$ is the diameter of the largest fragment allowed by the distribution. A power-law distribution accounts for the demonstrated fractal nature of high energy fragmentation events \citep{Turcotte2018...91(B2)...1921} which holds strongly for meteoric entries \citep{2021ICARUS...367...x, 2013GeocI...57...583, 2014M&PS...49...1989}.
\subsection{Ablation Model}

Classical meteor theory provides equations governing the deceleration and ablation of the fragments produced during meteoric breakup \citep{McKinley(1961),1980Icarus...42...211,Bronshten(1983),1997JRASC...91...x,Trigo2021}, 

\begin{equation}
\frac{\mathrm{d} v}{\mathrm{~d} t}=-\frac{\Gamma \rho_{\mathrm{a}} v^2}{m} A\left(\frac{m}{\rho_{\mathrm{m}}}\right)^{2 / 3},
\end{equation}
\begin{equation}
\frac{\mathrm{d} m}{\mathrm{~d} t}=-\frac{\Lambda}{2 \zeta} A\left(\frac{m}{\rho_m}\right)^{2 / 3} \rho_a v^3,
\end{equation}
where $v$ is the instantaneous atmospheric fragment speed, $m$ is the instantaneous fragment mass, $\Gamma$ is a dimensionless drag coefficient, $\Lambda$ is a dimensionless heat transfer coefficient, $\zeta$ is the heat of ablation (defined as the summed heats of fusion and vaporization), $\rho_m$ is the material density of the impactor, and $\rho_a$ is the atmospheric density. ``\textit{A}'' is a shape factor defined such that $A\left(m/\rho_m\right)^{2 / 3}$ equals the cross sectional area of a given fragment. Assuming a spherical geometry \citep{1980Icarus...42...211, 2005M&PS...40...817, 2021ICARUS...367...x}, the value of ``\textit{A}'' follows directly from solving for the cross-sectional area in terms of fragment volume. We define $\Omega$ to be the fragment's cross sectional area. Then, $\Omega=\pi\left(3/4\pi\right)^{(2 / 3)}\left(m/\rho_m\right)^{(2 / 3)}=1.21\left(m/\rho_m\right)^{(2 / 3)}$. For a sphere, $A = 1.21$. 

The trajectory of the bulk fragment cloud is given by, 
\begin{equation}
\frac{d z}{d t}=-v \sin (\gamma),
\end{equation}
\begin{equation}
\frac{d x}{d t}= v\sin (\lambda)\cos(\gamma),
\end{equation}
\begin{equation}
\frac{d y}{d t}= v\cos (\lambda)\cos(\gamma),
\end{equation}
where $z$ is the altitude, $\gamma$ is the angular trajectory relative to the ground, $\lambda$ is the azimuth, $x$ is the position along the East-West axis relative to the airburst location, and $y$ is the position along the North-South axis relative to the airburst location. Lastly, the atmospheric density profile is given by, $\rho_a=\rho_0 e^{(-z / H)}$, where $\rho_{0}$ is the sea-level atmospheric density, and $H = 8$ km is the scale height of the atmosphere \citep{2005M&PS...40...817}.

\section{IM1} 

The meteor IM1 was detected traveling at $v_{IM1} = 44.8$ $\mathrm{km \; s^{-1}}$ with $\gamma$ = 26.8$^{\circ}$ and $\lambda$ = 285.6$^{\circ}$ \citep{Zuluaga2019...3...68}. The total mass of the impactor was calculated as approximately $\mathrm{M} \sim 5 \times 10^5 \mathrm{~g}$ \citep{SirajarXiv...1904...07224}. Three distinct flares are apparent in the meteor’s light curve between altitudes 23 km and 18.7 km, with the largest flare occurring at 18.7 km which is taken to be the “gross” fragmentation altitude \citep{1993A&A...279...615}. Further analysis of the light curve provides the conservative yield strength of the meteor, $ Y_{IM1} \sim $113 MPa \citep{SirajRNAAS...6...81}, which is more than twice the yield strength of iron meteorites as calculated by \citet{SirajRNAAS...6...81}

\section{Fragmentation, Ablation, and Fragment Trajectory Assuming an Iron Composition}
The strongest known class of meteorites is iron \citep{JMS2001...36...1579}. Therefore, to provide a heuristic baseline estimate of the post-ablation fragment mass distribution of IM1, we initially adopt the corresponding material properties of iron. 
\subsection{Fragmentation of IM1}
The density of iron is $\rho_{\mathrm{m}}=7.8\mathrm{~g} \mathrm{~cm}^{-3}$. The implied diameter of the meteor, assuming a spherical geometry, is $D_{IM1} \approx 50$ cm. We choose $D_{\min} = 0.1$ cm, the smallest fragment diameter allowed by the empirical bounds of the NASA SBM \citep{2011OrbitalDebris...15...1}, and $D_{\max} = 0.7D_{IM1}$, as the upper limit of the distribution following \citet{2021ICARUS...367...x}. 
From equation (1), the pre-ablation fragment mass distribution function for IM1 is then, 

\begin{equation}
p_{m_{initial},IM1}=\frac{f}{3 \left(0.1^{-f}-0.7D_{IM1}^{-f}\right)}\left(\frac{7.8\pi}{6} \right)^{f / 3} m_{initial}^{-f / 3-1}
\end{equation}

\subsection{Ablation and Trajectory of IM1 Fragments} 

\begin{figure}[H]
\includegraphics[width=14.0 cm]{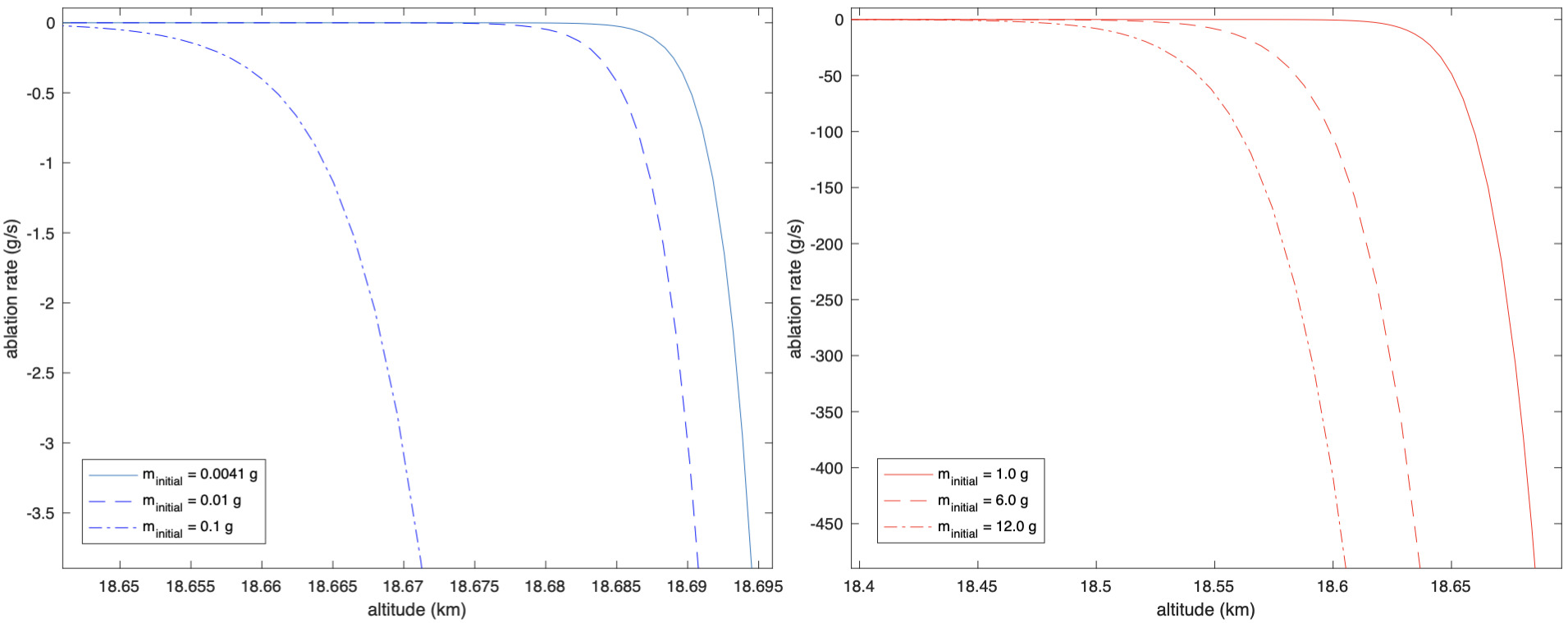}
\caption{Altitude vs ablation rate considering a range of $m_{initial}$ values which account for 99\% of the pre-ablation fragment mass distribution described in equation (7). Ablation rate is expressed as negative following from the definition of equation (3). (\textbf{a}) Left: $m_{initial} = .0041$ g, $m_{initial} = .01$ g, and $m_{initial} = .1$ g.(\textbf{b}) Right: $m_{initial} = 1.0$ g, $m_{initial} = 6.0$ g, and $m_{initial} = 12.0$ g.\label{fig1}}
\end{figure} 

The heat of ablation of iron is $\zeta=6.549 \times 10^{10} \mathrm{ \; erg} \;  \mathrm{g}^{-1}$. Previous studies on the ablation of iron meteors adopt a constant value for the dimensionless heat transfer coefficient, $\Lambda = .02$ \citep{1980Icarus...42...211}. Since the ablation rate is highly dependent on $\Lambda$, we conservatively use double this value, $\Lambda = .04$. For spherical bodies, the dimensionless drag coefficient is, $\Gamma = .5$ \citep{1980Icarus...42...211,1997JRASC...91...x}.

To calculate the ablation and surface impact location for a given pre-ablation fragment mass, $m_{initial}$, we integrate equations (2) through (6) starting at  $m_{initial},\; v_{IM1}$, the gross fragmentation altitude $z_0 = 18.7$ km, and the fragmentation site $(x = 0 \; \mathrm{km}, y = 0 \; \mathrm{km})$, along the angular trajectory of the bulk fragment cloud  $(\gamma = 26.8^{\circ}, \; \lambda = 285.6^{\circ}$). The integration is computed with a variable step, variable order (VSVO) integrator optimized for stiff differential systems. At each step we check that the fragment has not decelerated past its terminal velocity, and reset the velocity to its terminal value if this condition is not met. 

The ablative process is rapid. For $m_{initial} \in[.0041 \mathrm{~g}, 12 \mathrm{~g}]$, which accounts for 99\% of the distribution described in equation (7) (as shown in equation (8)), ablation concludes at approximately $z_0$ (Figure. 1). 

\begin{equation}
\int_{.0041}^{12.0} \left(\frac{f}{3 \left(0.1^{-f}-.7D_{IM1}^{-f}\right)}\left(\frac{7.8\pi}{6} \right)^{f / 3} m_{initial}^{-f / 3-1} \right) d m_{initial} = .99
\end{equation}

\subsection{Results and Analysis}

We define the probability distribution of the post-ablation fragment mass to be $p_{m_{final}}$. The distribution was constructed numerically via a Monte Carlo Simulation. $5 \times 10^5$ $m_{initial}$ values were randomly sampled from equation (7). The post-ablation mass, defined as $m_{final}$, and the surface impact site corresponding to each $m_{initial}$ value were calculated in accordance with section 4.2. The probability distribution function of the post-ablation fragment masses was then computed as a normalized histogram (Figure. 2). 

\begin{figure}[H]
\includegraphics[width=14.0 cm]{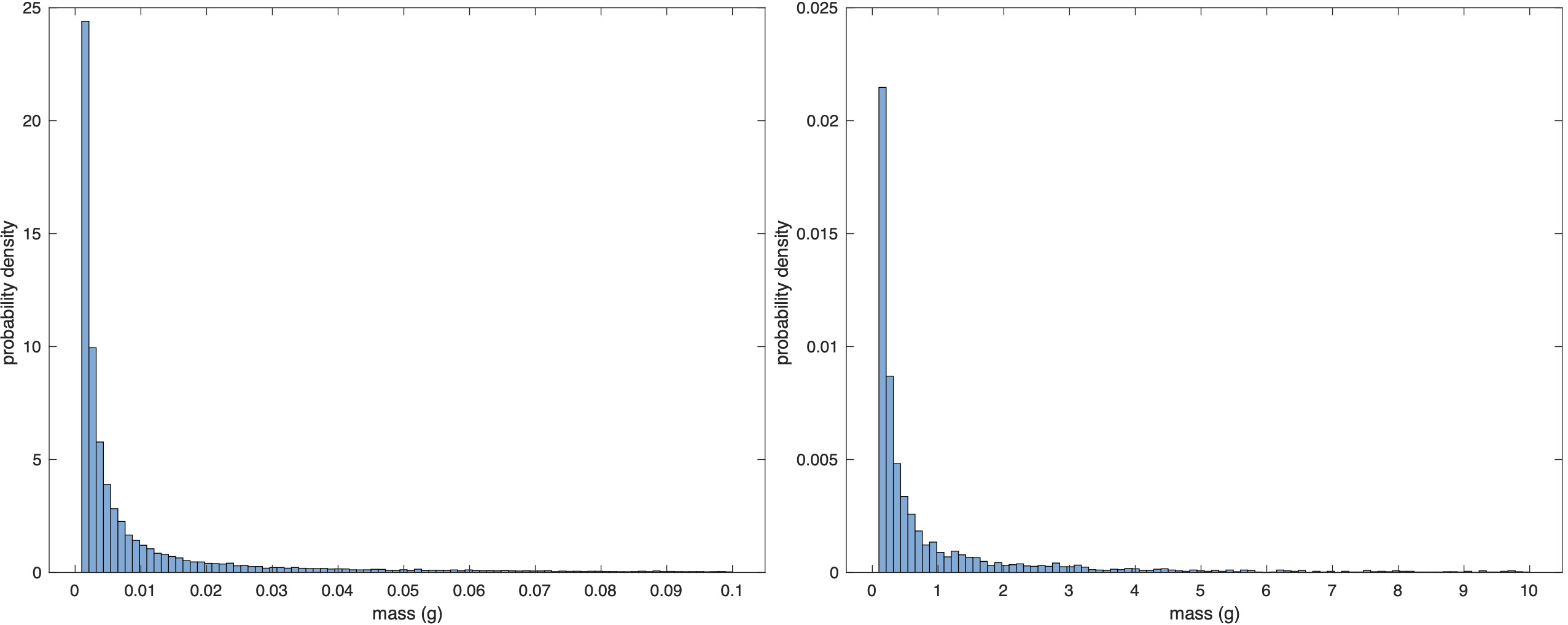}
\caption{Post-ablation fragment mass probability density histogram over the regions,(\textbf{a}) Left: $m_{final} \in[.001\mathrm{~g}, .1\mathrm{~g}]$. (\textbf{b}) Right: $m_{final} \in[.1\mathrm{~g}, 10\mathrm{~g}]$}.\label{fig1}
\end{figure} 

\begin{figure}[H]
\centering
\includegraphics[width=10.0 cm]{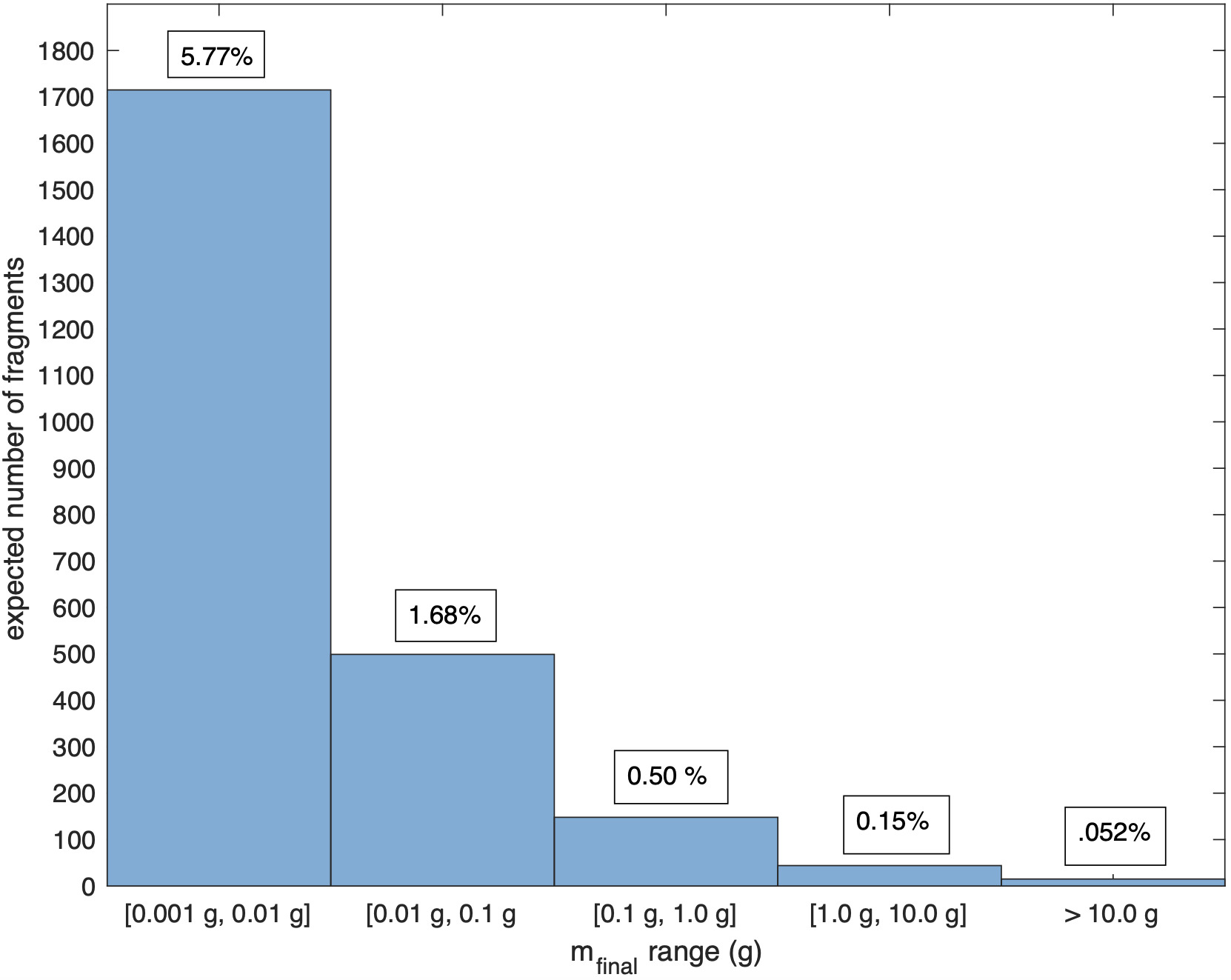}
\caption{Percentage of iron fragments surviving with a post-ablation mass in the given ranges and the expected number of fragments in each range}
\end{figure} 

As expected from empirical studies on recovered meteor fragments \citep{2013GeocI...57...583, 2014M&PS...49...1989}, the distribution $p_{m_{final}}$ has the structure of a power law. Summing the area under the normalized histogram for $m_{final} \geq .001$ g, we find that 8.14\% of IM1 fragments survive ablation with a mass $m_{final} \geq 0.001$ g.

 From section 3, the total mass of IM1 was calculated to be, $\mathrm{M} \sim 5 \times 10^5 \mathrm{~g}$ \citep{SirajarXiv...1904...07224}. Following \citet{2021ICARUS...367...x}, the total number of fragments produced during the airburst is calculated as, 

\begin{equation}
N=\frac{M}{\int_{m_{\min }}^{m_{\max }} m p_m d m} = 2.93 \times 10^4
\end{equation}
where $m_{min} = \rho_m(4/3)\pi(D_{min}/2)^3 = 4.10 \times 10^{-3}$ g and $m_{max} = \rho_m (4/3)\pi(D_{max}/2)^3 = 1.75 \times 10^5$ g. 
The expected number of fragments in five subdivisions of the post-ablation mass range $m_{final} \geq .001$ g as well as the specific probability of each subdivision is plotted in Figure. 3.

\subsection{Geographic Distribution} 

The trajectory of IM1's bulk fragment cloud followed a northwestern track over the ground ($\lambda$ = 285.6$^{\circ}$). In Figure. 4, we define the location of the IM1 airburst as the origin $(x = 0 \; \mathrm{km}, y = 0 \; \mathrm{km})$. We then plot the approximate surface impact site along the "line of highest probability" \citep{SpurnyAA...570...A39} drawn by the trajectory of IM1's bulk fragment cloud for post-ablation masses $m_{final} \in[0.001 \mathrm{~g}, 1.0 \mathrm{~g}]$. This mass range accounts for  97.5\% of post-ablation fragments with a mass $\geq$ .001 g. We include the normalized marginal histograms for the western and northern impact coordinates of the fragments.

\begin{figure}[H]
\centering
\includegraphics[width=14.0 cm]{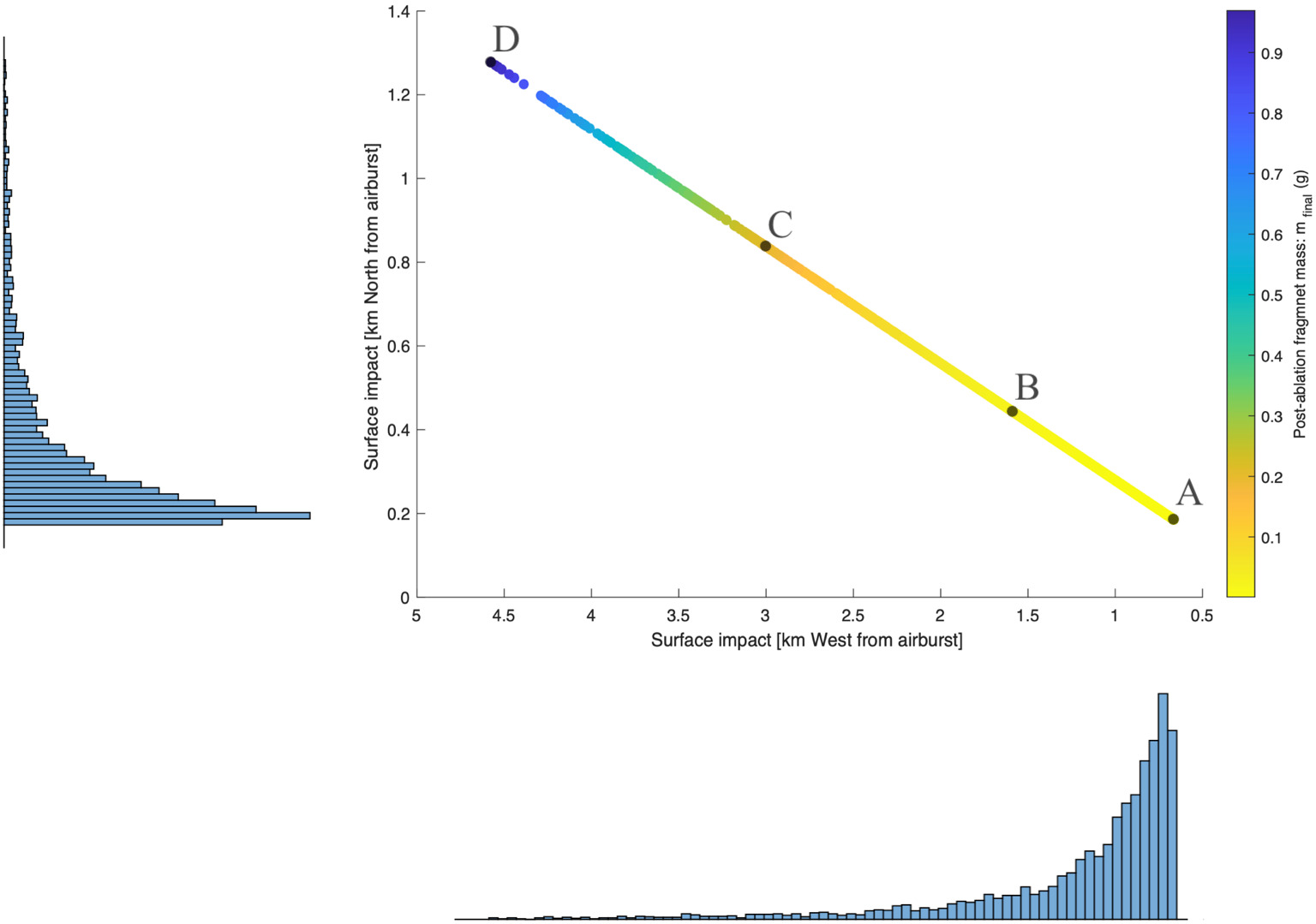}
\caption{Surface impact location and marginal histograms for the western and northern impact coordinates of post-ablation fragment masses, $m_{final} \in [0.001 \textrm{ g}, 1.0 \textrm{ g}]$. The "y" axis of the northern impact histogram corresponds to the y axis in the plot. The "x" axis of the western impact histogram corresponds to the x axis in the plot. The impact site is a strong function of mass, with larger post-ablation masses landing further from the fragmentation site.  }  
\end{figure} 

To estimate the geographic concentration of fragments, we sum the area under the northern impact coordinate histogram in subdivisions of the range $y \in[.19 \; \mathrm{km}, 1.28 \; \mathrm{km}]$ and the western impact coordinate histogram in subdivisions of the range $x \in[.66 \; \mathrm{km}, 4.58 \; \mathrm{km}]$. 82\% of considered post-ablation fragments ($\sim 1900$ fragments) are expected to fall approximately along the $\sim$1.0 km length between points ``A'' ($x = .66$ km, $y = .19$ km) and ``B'' ($x = 1.6$ km, $y = .44$ km) drawn out by the bulk trajectory of the IM1 fragment cloud. This region predominately contains fragments with a post-ablation mass $m_{final} \in[0.001 \mathrm{~g}, 0.1 \mathrm{~g}]$. 

14.6\% of considered post-ablation fragments with a mass $\geq .001 \mathrm{~g}$ ($\sim 340$ fragments) are expected to fall approximately along the 1.5 km length between points ``B'' and ``C'' ($x = 3.0$ km, $y = 0.84$ km). This region predominately contains fragments with a post-ablation mass  $m_{final} \in[0.1 \mathrm{~g}, 0.3 \mathrm{~g}]$. 

The remaining $\sim$ 3.4\% of considered post-ablation fragments with a mass $\geq .001 \mathrm{~g}$ ($\sim 79$ fragments) are estimated to fall approximately along the 1.66 km length between points ``C'' and ``D'' ($x = 4.60$ km, $y = 1.28$ km). This final region predominately contains fragments with a post-ablation mass, $m_{final} \in[0.3 \mathrm{~g}, 1.0 \mathrm{~g}]$.

\section{Additional Constraints on the 1M1 Assuming a Steel Composition}

As iron is the strongest known class of meteorites, the previous sections assumed an iron material for IM1 which has a yield strength calculated by \citet{SirajRNAAS...6...81} as $Y_{iron} = 50$ MPa. However, as previously stated, the calculated yield strength of IM1 is $Y_{IM1} \sim 113$ MPa. Therefore, to bracket the post-ablation fragment mass distribution in terms of the strength of the impactor, we now consider the post-ablation fragment mass distribution assuming a generic steel material, $Y_{steel} = 250$ MPa. 

\subsection{Fragmentation of IM1}

The density of steel is $\rho_{\mathrm{m_{steel}}}=8.0 \mathrm{~g} \mathrm{~cm}^{-3}$. Given the known mass $\mathrm{M} \sim 5 \times 10^5 \mathrm{~g}$, and again assuming a spherical impactor, the implied diameter is $D_{IM1_{steel}} \approx 50$ cm. With reference to equation (1), for the maximum fragment diameter produced during the airburst we again adopt $D_{\max_{steel}} = .7D_{IM1_{steel}}$. 

To estimate the minimum pre-ablation steel fragment diameter, we first scale the mass of the smallest pre-ablation steel fragment in proportion to the increase in yield strength of steel versus iron - a factor of five ($Y_{steel}/Y_{iron} = 250 \textrm{ MPa} / 50  \textrm{ MPa} = 5.0$). Recall the minimum pre-ablation fragment mass assuming an iron material was $m_{min} = .0041 $ g. Therefore, $m_{min,steel} = 5.0\times m_{min}$ = .0205 g. We then calculate the minimum pre-ablation steel fragment diameter as $D_{min,steel} = 2(3m_{min,steel}/4\pi\rho_{m_{steel}})^{(1/3)} \approx $ 0.17 cm. The pre-ablation fragment mass distribution assuming a generic steel material becomes, 

\begin{equation}
p_{m_{initial,steel}}=\frac{f}{3 \left(.17^{-f}-.7D_{IM1_{steel} }^{-f}\right)}\left(\frac{8\pi}{6}\right)^{f / 3} m_{initial,steel}^{-f / 3-1}.
\end{equation}

\subsection{Ablation of IM1 Fragments}

The thermal properties of steel are comparable to thermal properties of iron. The heat of ablation of steel is $\zeta_{steel}=6.747 \times 10^{10} \mathrm{ \; erg} \;  \mathrm{g}^{-1}$ \citep{NuclearEngineering...57...74} and we again choose the dimensionless heat transfer coefficient as $\Lambda_{steel} = .04$. We randomly sample $5 \times 10^5$ $m_{initial,steel}$ values from equation (10) and calculate each fragment's post-ablation mass, defined as $m_{final,steel}$, in accordance with Section 4.2.  We then construct the derived post-ablation fragment mass distribution assuming a steel material, defined as $p_{m_{final,steel}}$, as a normalized histogram depicted in Figure. 5.

By summing the area under the normalized histogram for $m_{final,steel} \geq 0.001$ g we find that assuming a steel material, $\sim 20.98\%$ of fragments survive with a mass $ \geq 0.001$ g. This is in comparison to the $\sim 8.14 \%$ of fragments that are expected to survive with a mass $m_{final} \geq 0.001$ g when assuming an iron material. 

\begin{figure}[H]
\centering
\includegraphics[width=14.0 cm]{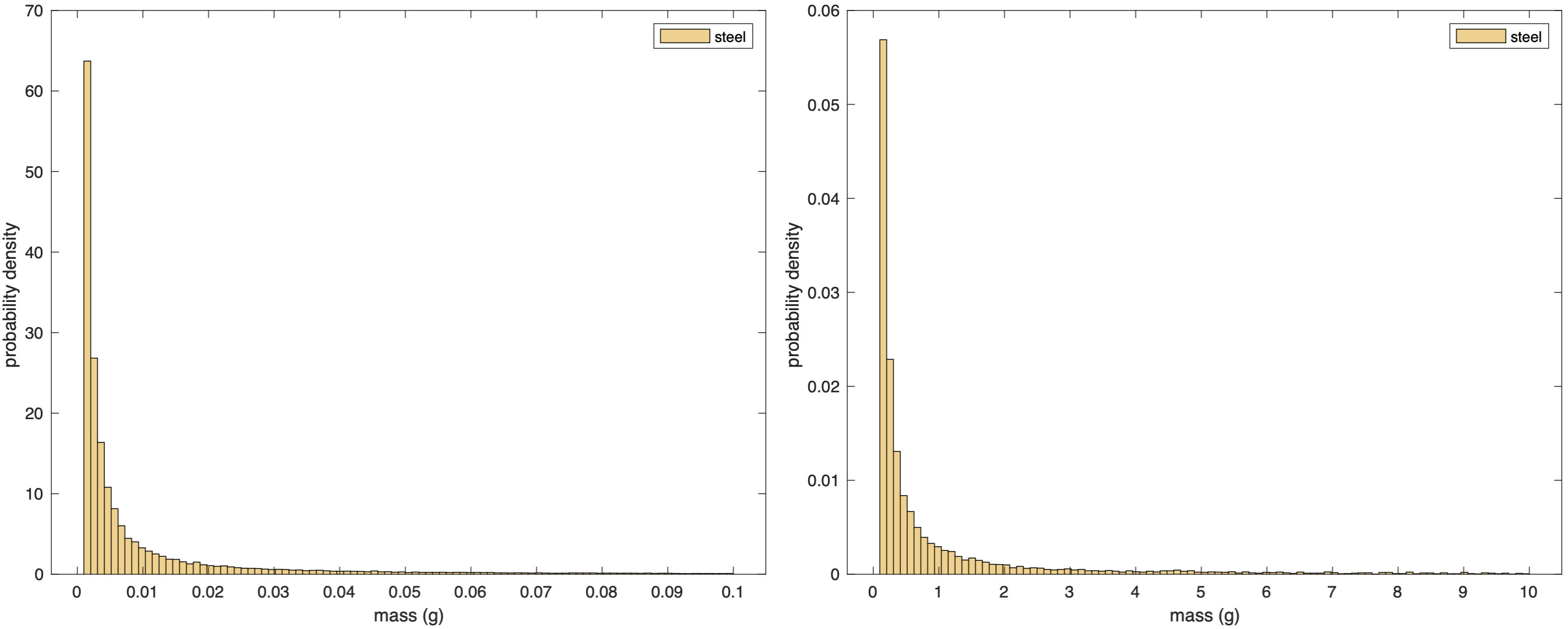}
\caption{Post-ablation fragment mass probability density histogram over the regions,(\textbf{a}) Left: $m_{final,steel} \in[.001\mathrm{~g}, .1\mathrm{~g}]$. (\textbf{b}) Right: $m_{final,steel} \in[.1\mathrm{~g}, 10\mathrm{~g}]$}.\label{fig1}
\end{figure} 

The total number of fragments produced during the airburst is calculated as, 

\begin{equation}
N=\frac{M}{\int_{m_{\min,steel }}^{m_{\max,steel }} m p_{m_{initial,steel}} d m} = 1.22 \times 10^4, 
\end{equation}
where $m_{min,steel}$ is as previously calculated and $m_{max,steel} = \rho_{m,steel} (4/3)\pi(D_{max,steel}/2)^3 = 1.79 \times 10^5$ g. From equation (11), we see that because of steel's higher yield strength in comparison to iron, fewer steel fragments are expected to be produced during the airburst. However, the fragments produced have a greater mass from the range $[2.05 \times 10^{-2} \; \mathrm{g}, \; 1.79 \times 10^5 \; \mathrm{g}]$.

We calculate the expected number of fragments in five subdivisions of the post-ablation mass range 
$m_{final,steel} \geq 0.001$ g as well as the specific probability of each subdivision and plot the results in Figure. 6 below.

\begin{figure}[H]
\centering
\includegraphics[width=10.0 cm]{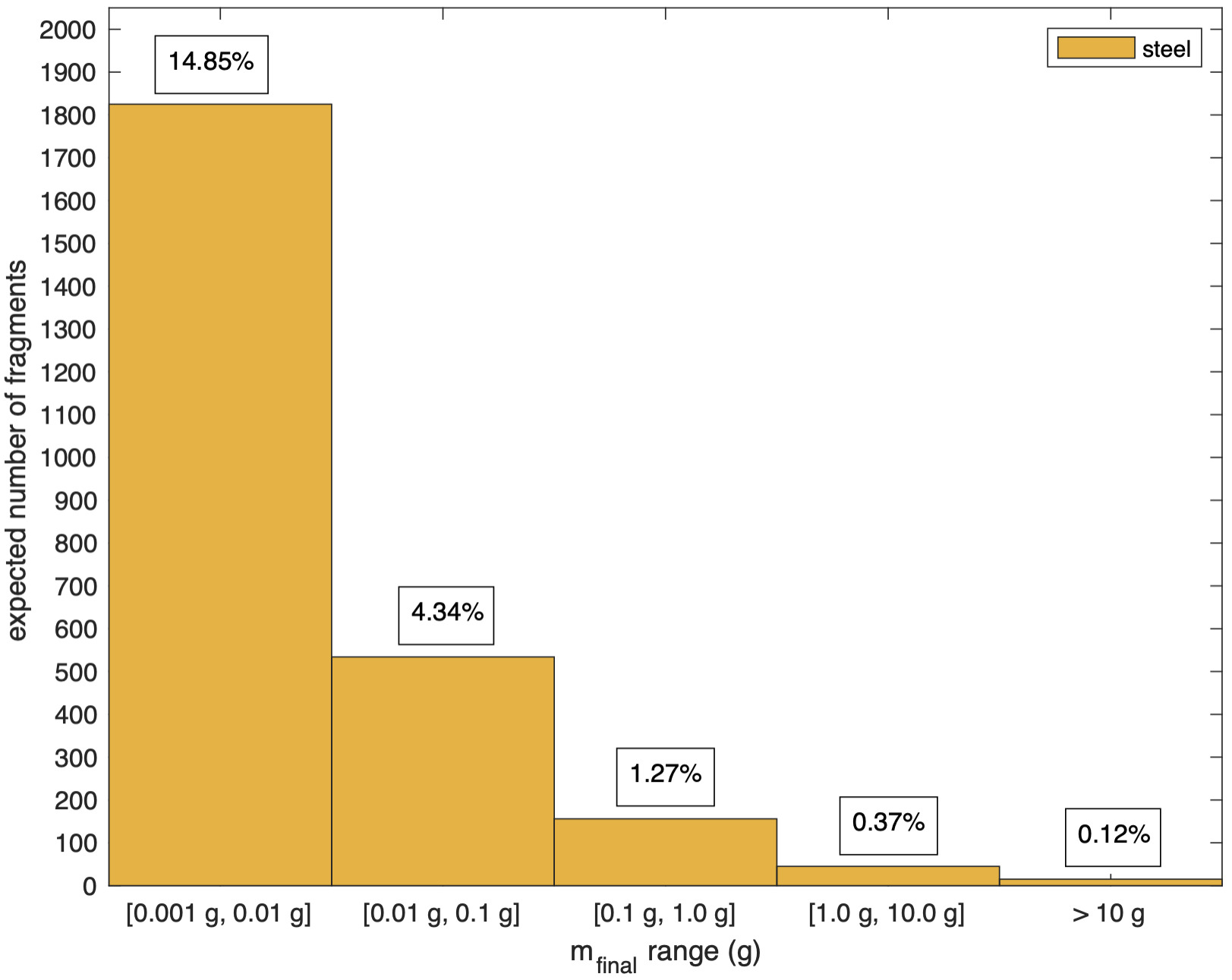}
\caption{Post-ablation fragment mass probability density histogram over the regions,(\textbf{a}) Left: $m_{final,steel} \in[.001\mathrm{~g}, .1\mathrm{~g}]$. (\textbf{b}) Right: $m_{final_steel} \in[.1\mathrm{~g}, 10\mathrm{~g}]$}.\label{fig1}
\end{figure} 

As shown in Figure. 6, the percentages of fragments with a post-ablation mass in each subdivision are higher than the corresponding percentages assuming an iron material. However, because fewer fragments are expected when assuming a steel impactor material, the overall number of fragments in a given post-ablation mass range is approximately equal between the iron and steel material analysis.

\subsection{Geographic Distribution}

\begin{figure}[H]
\centering
\includegraphics[width=14.0 cm]{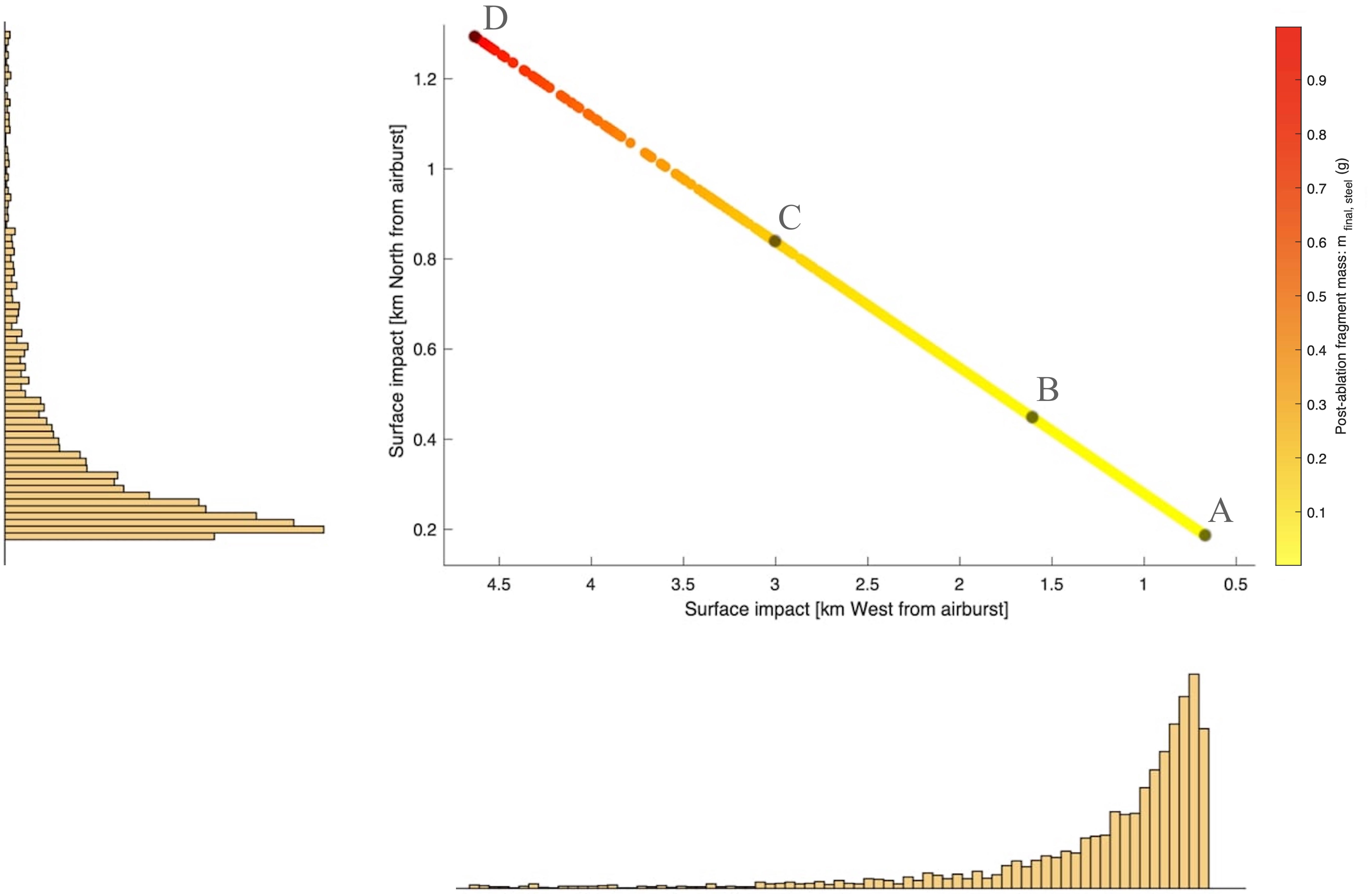}
\caption{Surface impact location and marginal histograms for the western and northern impact coordinates of post-ablation fragment masses, $m_{final,steel} \in [0.001 \textrm{ g}, 1.0 \textrm{ g}]$. The "y" axis of the northern impact histogram corresponds to the y axis in the plot. The "x" axis of the western impact histogram corresponds to the x axis in the plot. The impact site is a strong function of mass, with larger post-ablation masses landing further from the fragmentation site.}  
\end{figure} 

In line with Section 4.4, the surface impact site of IM1's post-ablation fragments assuming a steel material is a strong function of the fragments' final mass. The least massive fragments, $m_{final,steel} \in [0.001 \textrm{ g}, 0.1 \textrm{ g}]$, fall along the $\sim$ 1.0 km length between points "A" and "B" - almost directly underneath the airburst. Larger fragments with a mass > 0.1 g are expected to be distributed approximately along the line of highest probability (Figure. 7, Segment "B-C" and Figure. 7, Segment "C-D"). 

Note that because the surface impact site is dependent on the fragments' post-ablation mass, and because the expected number of fragments in a given post-ablation mass range are similar between the iron and steel analysis (Figure. 3 and Figure 6), the surface impact distributions assuming iron and steel materials are also similar. In Figure. 7, $\sim$ 2000 fragments are expected to fall between points "A" and "B", predominately with a mass $m_{final,steel} \in [0.001 \textrm{ g}, 0.1 \textrm{ g}]$. We expect hundreds of fragments between points "B" and "C", and tens of fragments between points "C" and "D". 

While it is true the varying density of distinct material compositions for IM1 would affect the fragments' volume, the difference in density between iron and steel is negligible. In addition, the steel and iron fragments are modeled as sharing the same spherical geometry. Therefore, neither density nor shape is a relevant factor in the comparative surface impact analysis between iron and steel compositions. 

\section{Conclusion}

We find that even for an impactor speed as high as 44.8 $\; \mathrm{km \; s^{-1}}$, a significant portion of meteor fragments survive ablation. In the case of IM1, we estimate assuming an iron material that 8.14\% of IM1 fragments survive ablation with a final mass $\geq$ .001 g. 

Of this 8.14\% of fragments, 70.9\% (1700 fragments) have a mass $\in[.001 \mathrm{~g}, .01 \mathrm{~g}]$. The majority of this fragment group is expected to fall approximately along a 1.0 km length immediately to the northwest of the airburst (Figure. 4, Segment ``A-B''). The optimal search area for small fragment recovery is therefore approximately under the airburst site. Larger fragments are expected to be distributed approximately along the line of highest probability drawn by the bulk trajectory of the IM1 fragment cloud, at a distance greater than 
$\sim$ .44 km north and $\sim$ 1.60 km west (Figure. 4, Segment ``B-C'' and Figure. 4, Segment ``C-D''). 

 To bracket the post-ablation fragment mass distribution based on material strength, we calculate the post-ablation fragment mass distribution based on a steel material. Assuming the properties of steel, we estimate 20.98\% of fragments survive ablation with a post-ablation mass $\geq$ .001 g. Of these fragments, the majority are again expected to land approximately underneath the airburst site.

\authorcontributions{Conceptualization: A.L. and A.S.; methodology: A.L., A.S, A.T.R; software: A.T.R, A.S; formal analysis: A.T.R, A.S; resources, A.L.; data curation: A.T.R; writing---original draft preparation: A.T.R; writing---review and editing: A.L, A.S; visualization: A.T.R, A.S; supervision: A.L.; project administration: A.L.; funding acquisition, A.L. All authors have read and agreed to the published version of the manuscript.}

\funding{This project was supported in part by the Galileo Project at Harvard University.}

\dataavailability{Data sets are available from the authors upon request.} 

\conflictsofinterest{The authors declare no conflict of interest.} 


\abbreviations{Symbol Definitions}{
The following symbols are used in this manuscript:\\

\noindent 
\begin{tabular}{@{}ll}
$D_{min}$ & Minimum pre-ablation fragment diameter (iron)\\
$D_{max}$ & Maximum pre-ablation fragment diameter (iron)\\
$\rho_{m}$ & Impactor density (iron)\\
$D_{min_{steel}}$ & Minimum pre-ablation fragment diameter (steel)\\
$D_{max_{steel}}$ & Maximum pre-ablation fragment diameter (steel)\\
$\rho_{m_{steel}}$ & Impactor density (steel)\\
$f$  & Scale factor\\
$\rho_a$  & Atmospheric density\\
$\rho_0$  & Sea level atmospheric density\\
$v$ & Instantaneous atmosperhic fragment speed\\
$m_{initial}$ & Pre-ablation fragment mass (iron)\\
$m_{initial,steel}$ & Pre-ablation fragment mass (steel)\\
$m$ & Instantaneous atmosperhic fragment mass\\
$m_{final}$ & Post-ablation fragmnet mass (iron)\\
$m_{final,steel}$ & Postµ-ablation fragment mass (steel)\\
$H$ & Atmospheric scale height\\
$\Gamma$ & Dimensionless drag coefficient\\
$\Lambda$ & Dimensionless heat transfer coefficient\\
$\zeta$ & Heat of ablation\\
$\gamma$ & Fragment cloud's angular trajectory relative to the ground\\
$\lambda$ & Fragment cloud's azimuth\\
\end{tabular}
}


\begin{adjustwidth}{-\extralength}{0cm}

\reftitle{References}

\PublishersNote{}
\end{adjustwidth}
\end{document}